\documentstyle[psfig]{mn}

\title{The origin of the high velocity circumstellar gas around
SN~1998S}
 
\author[N.N.~Chugai, S.I.~Blinnikov, 
A.~Fassia, P.~Lundqvist, W.P.S.~Meikle and E.I.~Sorokina]
{N.N.~Chugai$^1$, S.I.~Blinnikov$^2$, A.~Fassia$^3$, P.~Lundqvist$^4$, 
  W.P.S.~Meikle$^3$,
 \newauthor  and E.I.~Sorokina$^5$\\
$^1$Institute of Astronomy, RAS, Pyatnitskaya 48, 109017 Moscow,
 Russia\\
$^2$ITEP, 117218, Moscow, Russia\\
$^3$Astrophysics Group, Blackett Laboratory, Imperial College, 
 Prince Consort Rd. London SW7 2BZ, UK\\
$^4$Stockholm Center for Physics, Astronomy and Biotechnology, 
Stockholm Observatory, Department of Astronomy,\\SE-10691 Stockholm, Sweden\\
$^5$Sternberg Astronomical Institute, 119899 Moscow, Russia\\
}

\date{Accepted 2001
      Received 2000;
      in original form 2000}

\pagerange{\pageref{firstpage}--\pageref{lastpage}}
\pubyear{2001}

\begin{document}

\maketitle

\label{firstpage}

\begin{abstract}

Modelling of high resolution Balmer line profiles in the early-time
spectra of SN~1998S shows that the inferred fast ($\approx 400$ km
s$^{-1}$) circumstellar (CS) gas on days~23 and 42 post-explosion is
confined to a narrow, negative velocity gradient shell just above the
photosphere.  This gas may be identified with a slow ($v< 40$ km
s$^{-1}$) progenitor wind accelerated at the ejecta-wind interface.
In this scenario, the photosphere coincides with a cool dense shell
formed in the reverse shock.  Acceleration by radiation from the
supernova or by a shock-accelerated relativistic particle precursor
are both possible explanations for the observed fast CS gas.  An
alternative, equally plausible scenario is that the fast CS gas is
accelerated within shocked clouds engulfed by the outer shock, as it
propagates through the intercloud wind.

\end{abstract}


\section{Introduction}

Supernova (SN) 1998S in NGC~3877 was discovered by Li \& Wan (1998),
probably before maximum light (Fassia et al. 2000).  The presence of
narrow lines in its early-time spectra led to its classification as a
type~IIn (SN~IIn) (Leonard et al. 2000).  Nevertheless, the close
resemblance of its light curve and spectral development to those of
SN~1979C (Liu et al. 2000) means that it is also valid to classify
SN~1998S as another bright SN~IIL.  Whatever the assigned class, there
is strong evidence for the existence of a dense circumstellar (CS)
wind in its immediate vicinity of SN~1998S.  This includes the
presence of narrow CS lines (Bowen et al. 2000; Fassia et al. 2001),
broad emission lines originating from the interaction (Leonard et
al. 2000; Gerardy et al. 2000; Fassia et al. 2001), and radio and
X-ray emission (Pooley et al. 2001).  Despite extensive effort in the
past, we still have a poor understanding of the origin of the large
pre-supernova mass loss and the ejecta-wind interaction phenomenon in
SNe~IIn.  

SN~1998S is the most comprehensively observed SN~IIn, thus providing
us with an excellent opportunity for gaining a deeper insight into the
morphology and origin of the dense CS environment and the physics of
the interaction.  Of particular interest is the origin of two distinct
CS wind velocity components deduced from high resolution early-time
spectra of SN~1998S taken at $\sim$2--5 weeks after the explosion
(Bowen et al. 2000; Fassia et al. 2001). These had velocities of
$u_{\rm w}=40-50$ km s$^{-1}$, and $u\approx 400$ km s$^{-1}$,
respectively.  The slower component is revealed by both narrow H~I and
He~I P~Cygni profiles and forbidden lines, while the faster component
is seen only in broader H~I and He~I absorption lines.  The minima of
the broad H$\alpha$ and H$\beta$ absorptions shifted from
$\sim$--115~km s$^{-1}$ on 1998 March~20 to $\sim$--85~km s$^{-1}$ by
1998 April~8.  Similarly broad CS absorption lines are also seen in
high resolution ultraviolet (UV) spectra via the Mg~II 2800~\AA\
doublet (Bowen et al. 2000).  While the presence of the slow wind is
unsurprising for a SN~II progenitor, the origin of the faster CS
material (CSM) is not clear.  Bowen et al. (2000) suggest it
originated as a pre-existing fast wind component (fast shell).  Fassia
et al. (2001) suggest it was due to a faster wind produced during a
recent BSG phase. A third possibility is that it was due to pre-shock
radiative acceleration of the slow wind (Lentz et al. 2001; Fassia et
al. 2001).

In this paper we shall focus on the nature of the fast CS component
which, we believe, is related to the structure of both the wind and
the ejecta-wind interface.  We start with a study of the kinematics
and location of the fast CS gas (Section 2), making use of the high
resolution profiles of Balmer lines observed on 1998 March 20 and 1998
April 8 (Bowen et al. 2000; Fassia et al. 2001).  Through our
modelling of the CS gas line profiles, we reveal that the fast CS
component appears to lie very close to the SN photosphere.  Indeed,
this proximity presents a challenge in understanding the structure of
the ejecta-wind interface.  A solution will be proposed.  We then
consider the plausibility that the fast CS gas results from the
pre-shock acceleration of the slow wind by radiation or relativistic
particles from the SN.  (Section 3).  As an alternative explanation,
we also investigate shocked dense wind clumps (Section 4).  We note
that the CS wind considered here corresponds to the
`Mid-CSM/Outer-CSM' regions defined in Fassia et al. (2001).  This
will be discussed further in Section~5, where we conclude that the
division into Mid- and Outer-CSM zones is probably not required after
all.  The `Inner-CSM' region (Fassia et al. 2001), overrun by the
ejecta around 2~weeks post-explosion, is examined by Chugai (2001).

Based on Chugai's (2001) model for the SN-CSM interaction and the
light curve, we prefer 1998 February 25 as the date of the SN~1998S
explosion.  Adopting this as zero epoch, March 20 and April 8 become
days~23 and 42, respectively.  This should be kept in mind when
comparing this work with other papers on SN~1998S.

\section{Modelling the circumstellar line profiles}

In the standard picture of spherically symmetric SN ejecta interaction
with a smooth CS wind, the ejecta-wind interface has a two-shock
structure with the outer and inner shocks running outward into the
wind and inward into the ejecta, respectively (Chevalier 1982;
Nadyozhin 1981, 1985).  We define a wind density parameter (linear
density) $w=4\pi r^2 \rho$, (g cm$^{-1}$), where $\rho$ is the density
of the unshocked CS wind.  With $w\sim 10^{16}$ g cm$^{-1}$, which is
typical for SNe~II with dense winds (e.g. SN~1979C, see Lundqvist \&
Fransson 1988), the inner shock wave is radiative.  Consequently, a
cool ($T\approx 10^4$ K) dense shell (CDS) of shocked SN ejecta forms
between the outer and inner shocks (Chevalier \& Fransson 1985, 1994).

To investigate the origin of the CS line profiles, we consider a model
of the line-forming region specified in terms of a) the photospheric
radius, $R_{\rm p}$, b) the extent of a shell of fast CS gas,
$R_1<r<R_2$, and c) the extent of a slow external wind, $R_2<r<R_3$.
We express the radii in terms of the photospheric radius, i.e. $R_{\rm
p}=1$, unless the physical units are needed.  The model is similar, in
some respects, to that of Lentz et al. (2001), but with two important
differences.  First, in their model there is a significant ``SN
atmosphere'' zone lying between the SN photosphere and fast CS gas
($R_{\rm p}<r<R_1$).  In our model the SN atmosphere is absent (we
explain why below).  There may be a layer of very hot ionised
post-shock gas between the photosphere and the fast CS gas ($R_{\rm
p}<r<R_1$), which does not contribute to line emission or absorption
in the optical, but reprocesses the photospheric spectrum in the
ultraviolet due to Comptonization (Lundqvist et al., in preparation).
Second, in Lentz et al. the velocity within the fast CS gas is assumed
constant, while in our model the velocity can vary with radius.  Since
the kinematic structure of the resulting fast CS gas is not known, we
consider initially three different possibilities: (a) a fast, constant
velocity wind, (b) an accelerated flow with a positive velocity
gradient, $v\propto r$, (this might arise from a CS shell lost in a
violent mass ejection before the SN explosion (Grasberg \& Nadyozhin
1987)), (c) an accelerated flow with a negative velocity gradient
($dv/dr<0$), characteristic of pre-shock acceleration of the slow
wind.  For the case with a negative velocity gradient, we adopt as the
description of the fast CS gas velocity $u$:

\begin{equation}
u=(u_{\rm f}-u_{\rm w})\left[1-
\left(\frac{r-R_1}{R_2-R_1}\right)^{\omega}\right] + u_{\rm w},
\label{eq:vel}
\end{equation}

\noindent where $u_{\rm w}$, the velocity of the slow wind, is assumed
to be constant with radius.  Thus the velocity of the accelerated CS
gas is $u_{\rm f}$ at its innermost limit $R_1$, falling to $u_{\rm
w}$ at its outermost limit $R_2$ where it merges with the
unaccelerated CS wind.  Note that $\omega\ll 1$ corresponds to a very
steep velocity gradient, while $\omega\gg 1$ gives a mild slope.

\begin{figure}
\centerline{\hspace{0.0cm}
\psfig{file=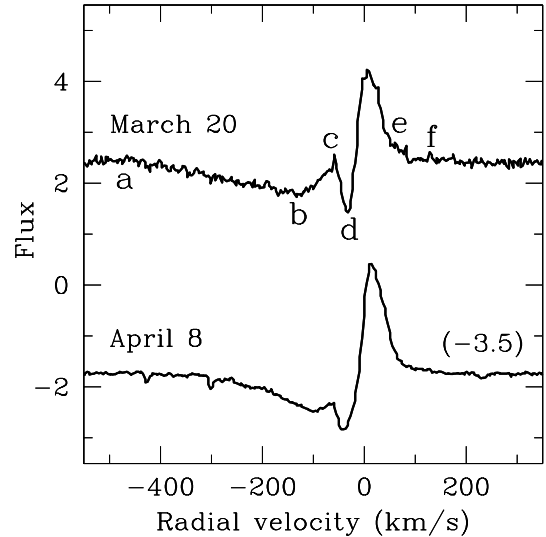,height=10cm,angle=0}
}
\caption{High resolution H$\alpha$ spectra of SN~1998S for two epochs
(Fassia et al. 2001).  The flux is in units of $10^{-14}$ erg cm$^{-2}$
s$^{-1}$ \AA$^{-1}$. The lower spectrum is shifted down by the value
shown in parentheses.  The symbols a--e show features referred to in
the text. }
\label{fig1}
\end{figure}

Given the population of the second level $n_2(r)$, and the line source
function $S(r)$, the CS gas line profile can then be easily calculated
for the velocity fields specified in (a)--(c) above.  We do not
compute the actual hydrogen level population in the wind (although
this is itself an interesting problem).  Instead, we parameterize the
behaviour of $n_2(r)$ and $S(r)$.  Specifically, we adopt a power law
distribution for the second level population $n_2(r)\propto r^{-p}$
for $r>R_1$, with $n_2(R_1)$ and $p$ as free parameters.  We split the
H$\alpha$ source function into net emission and scattering terms,
respectively $S_{\rm e}$ and $S_{\rm sc}$, with $S_{\rm e}\propto
r^{-q}$ and $S_{\rm sc}=WI_{\rm c}$, where $W$ is the dilution factor
and $I_{\rm c}$ is the continuum intensity.  This representation
reduces uncertainty in the H$\alpha$ source function.  For other
Balmer lines we cannot assume such a simple representation of the
source function.  However, the scattering term contribution is
expected to be substantially lower than for H$\alpha$, and so we
simply adopt the power law $S\propto r^{-q}$ for these lines.  Rather
than using the Sobolev escape approximation, we directly integrated
the radiative transfer equation, adopting a Gaussian local profile
with a thermal most probable velocity $u_{\rm th}=\sqrt{2kT/m}$.  The
model parameters comprised $R_1$, $R_2$, $R_3$, $u_{\rm f}$, $u_{\rm
w}$, $u_{\rm th}$, $\omega$, $p$, $q$, $n_2(R_1)$ and $S(R_1)$.

\begin{table}
  \caption{Parameters of illustrative fast CS gas line profile models}
  \bigskip
  \begin{tabular}{ccccccc}
  \hline
 No. & Kinematics  & $R_1$  &  $R_2$ & $\omega$ & $p$ & $q$ \\ 

\hline 
 
 1  & $v=\mbox{const.}$ &  1   &   2    &           &   3  &  6  \\
 2  & $v\propto r$      &  1   &   2    &           &   3  &  6  \\
 3  &  $dv/dr<0$        &  1   &  1.3   &    0.65   &   2  &  2 \\
 4  &  $dv/dr<0$        &  1   &  1.5   &    0.65   &   2  &  2 \\
 5  &  $dv/dr<0$        &  1.1 &  1.3   &    0.65   &   2  &  2 \\
 6  &  $dv/dr<0$        &  1   &  1.3   &    1.3    &   2  &  2 \\
 7  &  $dv/dr<0$        &  1   &  1.3   &    0.3    &   2  &  2 \\

\hline 
\end{tabular}
\label{tab1} 
\end{table}

\begin{figure}
\centerline{\hspace{0.0cm}
\psfig{file=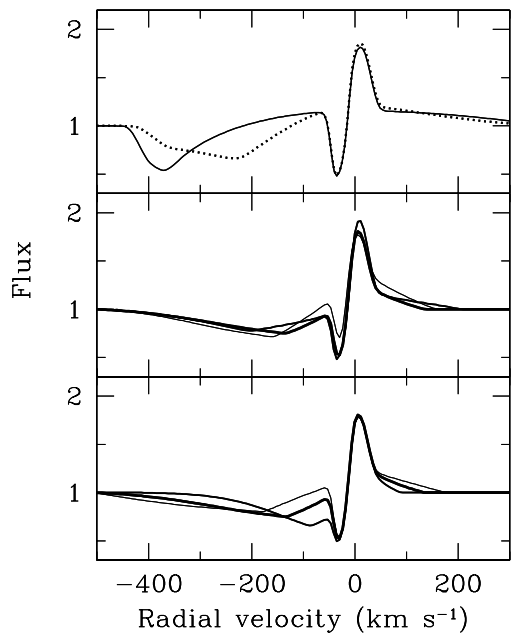,height=10cm,angle=0}
}
\caption{
Line profiles predicted by a two-component model consisting of an
inner fast flow and an outer slow wind.  The upper panel shows the
constant velocity fast wind Model~1 ({\em solid line}) and the
$v\propto r$ Model~2 ({\em dotted line}).  The middle and lower panels
show negative velocity gradient model profiles. The middle panel shows
the best-fit Model~3 ({\em thick line}), compared with Model~4 which
has a larger outer radius $R_2=1.5$ ({\em thinnest line}), and Model 5
which has a gap between the photosphere and the inner radius
($R_1=1.1$).  The lower panel also shows Model~3 ({\em thick line}),
compared here with Models~6 ({\em thinnest line}) and 7 which have,
respectively, steeper ($\omega=1.3$) and milder ($\omega=0.3$)
velocity gradients.
}
\label{fig2}
\end{figure} 

To assess the viability of the different CS kinematics suggested above
we shall concentrate on the kinematics of the fast flow, but always
including the slow wind component in the model.  In all the models the
maximum velocity of the fast flow, $u_{\rm f}=500$~km~s$^{-1}$.  The
kinematics of the slow wind is assumed to be a combination of its
steady outflow velocity $u_{\rm w}=30$~km s$^{-1}$ within which the
hydrogen gas has a thermal most probable velocity $u_{\rm th}=15$ km
s$^{-1}$ (corresponding to $\approx 14000$~K). The adopted values
for $u_{\rm w}$ and $u_{\rm th}$ are based on our best fits to the
observations for the slow wind lines (see below), although their
precise values are not crucial here.  A successful model of the fast
CS line component needs to be able to reproduce the specific features
indicated in Fig. 1, {\it viz. }  (i) the low velocity of the
absorption minimum (point b); (ii) the slowing down of the absorbing
gas velocity between day~23 and day~42, as indicated by the blue wing
(a--b); (iii) the relative weakness of the red emission wing (e--f),
presumably indicating an occultation effect.

The resulting model profiles are displayed in Fig. 2 with the
parameters given in Table 1.  Comparison of Figures 1 and 2 shows that
model 1 (fast, constant velocity wind) predicts a velocity for the
absorption minimum which is far greater than that observed.  Variation
of the free parameters ($R_1$, $R_2$, $p$, $q$) does not improve the
fit.  We therefore rule out this model.  A somewhat better match is
achieved with model 2 ($v\propto r$).  Moreover, if we increase
the outer radius by factor of three, then since we have a fixed
maximal velocity, the velocity of the absorption minimum can be made
even lower.  However, this model predicts that the characteristic
velocity derived from the fast CS gas line profile should increase
with time as SN overtakes the innermost, lower velocity CS gas.  In
fact we observe that the characteristic velocity {\it decreases} with
time (Fig.~1), and so we also discard this model.

Thus, we are left with the negative velocity gradient class of models
(Models 3--7). In Fig.~2 (middle panel) we show Model 3 which gives a
good fit to the observed day~23 profile (see below).  With Models
4--7, we illustrate the sensitivity of these negative gradient models
to parameter variations.  Increasing the outer limit, $R_2$ of the
accelerated CSM (Model 4) (Fig.~2 middle panel) produces a red wing
which is much stronger than that observed.  The introduction of a gap,
$\delta R=R_1-R_{\rm p}=0.1$, between the photosphere and the fast CS
gas (Model 5) also results in an unacceptably strong red wing (Fig.~2
middle panel).  Moreover, both Models 4 \& 5 exhibit significant
differences from Model 3 at radial velocities of $\sim-100$ km
s$^{-1}$.  Variation of the velocity gradient index, $\omega$, also
has a strong effect.  This is illustrated by models 6 \& 7, again
compared with Model 3 (Fig.~2, lower panel).  Thus, given a judicious
selection of parameters, these preliminary calculations demonstrate
that a model comprising a fast CS shell with a negative velocity
gradient can successfully reproduce the observations.

\begin{figure}
\centerline{\hspace{0.0cm}
\psfig{file=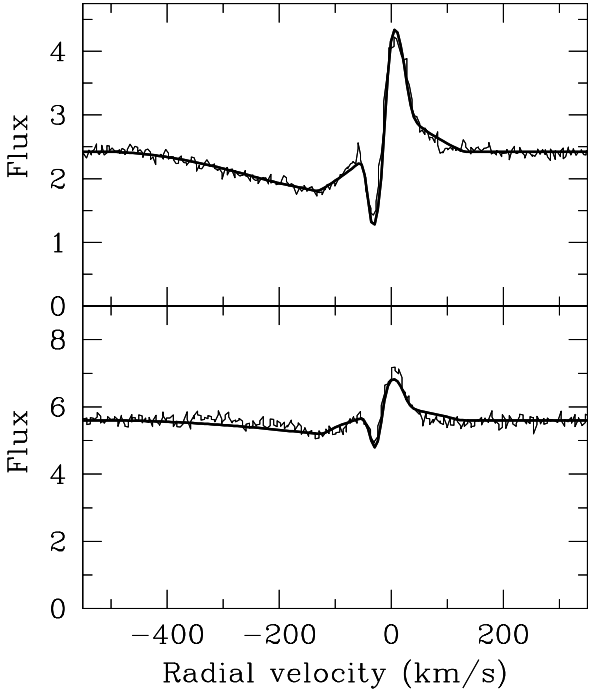,height=11cm,angle=0}
}
\caption{High resolution H$\alpha$ (top panel) and H$\beta$ spectral
profiles in SN~1998S on day~23. {\em Thin line} -- observation, {\em
thick line} -- model (Table 2) }
\label{fig3}
\end{figure}

We now fit the H$\alpha$ CS line profiles for both epochs and compute
the other Balmer line profiles using the same parameters as for
H$\alpha$, tuning only the amplitude of the source function.  A
redshift of +850 km s$^{-1}$ (Fassia et al. 2001) was adopted.  We
explored the model parameter space and found that the observed
H$\alpha$ profile could only be reproduced with a narrow (or zero) gap
between the photosphere and the fast CS gas, {\it viz.} $R_1 <
1.05R_{\rm p}$, with $R_1=R_{\rm p}$ being the best choice. The
power-law indices, $p$, $q$ and amplitudes, $n_2(R_1)$, $S(R_1)$ of,
respectively, the second level population and source function were
estimated by imposing the requirement that both the fast and slow CS
absorption components of all the hydrogen lines should be reproduced.
The slow wind outflow velocity $u_{\rm w}$ and thermal velocity
$u_{\rm th}$ were fixed by the positions and widths of the narrow
lines.  In general, the slow wind outer radius, $R_3$, is poorly
determined.  We tested values in the range $2.5<R_3<8$ but were always
able to produce acceptable fits by adjusting $p$ and $q$.  We have,
therefore, arbitrarily adopted $R_3=4$ on day~23 and $R_3=3$ on
day~42.  We judge that the Balmer line profile fits for days~23 and 42
(Figs.~3 \& 4) are quite good.  The parameters used are given in
Table~2.  In Fassia et al. (2001) we found that the narrow H-line
widths corresponded to a velocity of $\sim$60~km/s, and we suggested
that part of this was due thermal broadening. This is quantified here
by our modelling of the narrow hydrogen lines which indicates a slow
wind velocity of $30$ km s$^{-1}$ for hydrogen.  Curiously, this is
actually slightly slower than the $\sim$40~km/s wind velocity inferred
from the narrow forbidden lines (Fassia et al. 2001).

The model of the fast CS gas defined by the parameters in Table 2 is
henceforth dubbed the `kinematic model'.  The essential feature of
this model is that the fast CS gas is confined to a narrow layer,
$\Delta R/R\approx 0.3$, nearly attached to the photosphere.  This
indicates that the photosphere resides in the outermost layers of the
SN ejecta and possibly coincides with the CDS.  The latter point, in
fact, agrees with recent modelling of the SN~1998S interaction with
the CS gas, which demonstrates that the CDS remains opaque in the
Paschen continuum for about 45 days (Chugai 2001). The essential
elements of the kinematic model of the fast CS gas are illustrated in
Fig.~5.

This picture is, however, somewhat disturbing since it leaves no place
for the outer shock wave.  The point is that the fast CS gas
presumably lies in the pre-shock zone and therefore must be separated
from the CDS by the post-shock layer.  The extent of this layer
depends on whether the shock is adiabatic or radiative.  In the
adiabatic case the typical thickness of the postshock layer for the
outer shock is about $\Delta R/R \approx 0.2$ (Chevalier 1982).  This
is in conflict with the kinematic model of the pre-shock fast CS gas,
which requires $\Delta R/R < 0.05$.  On the other hand, if the outer
shock were essentially radiative, the post-shock layer might be thin
enough to meet this requirement.  However, the problem with the latter
possibility is that the wind density in the region $r>10^{15}$ cm,
where the interaction proceeds after March 16 (day 19) is very low, $w
\approx 10^{16}$ g cm$^{-1}$, according to the analysis of the fast CS
ultraviolet lines on March 16 (Lentz et al. 2001). (We note also, that
this is consistent with the lower limit of $\sim10^{14}$ g cm$^{-1}$ on
day~42 which can be obtained from the [O~III]-derived density lower
limit and the shock radius (Fassia et al.  2000, 2001)).  For such a
wind the outer shock is essentially adiabatic.  We therefore face a
`proximity problem', namely, how to reconcile the proximity of the
fast CS gas to the photosphere on days 23 and 42 with the claim that
the outer shock wave is adiabatic at these epochs.

Two solutions to this problem are conceivable.  The first possibility
is that the outer shock is adiabatic, {\em i.e} $R_1\approx 1.2$, but
that the CDS is strongly corrugated, presumably due to Rayleigh-Taylor
instabilities, in such a way that spikes of dense opaque CDS material
protrude into the postshock layer which lies in the region $R_{\rm
p}<r<R_1$.  The morphology of the corrugated CDS in this case is
similar to the `hedgehog' model used by Cid Fernandes \& Terlevich
(1994) to explain line profiles in SNe~II interacting with a dense CS
environment.  The thick forest of spikes can, in principle, produce an
efficient limb occultation of the far side of the fast CS shell and
thus bring about the effect of fast CS gas-photosphere proximity. 
  
The second solution is based upon the concept of a clumpy wind.  Wind
clumpiness is a rather common phenomenon for AGB and post-AGB stars
(Olofsson 1994) and is occasionally revealed in interacting SNe~II
(Chugai 1993).  We envisage a two-phase wind with dense clouds
embedded in the rarefied intercloud gas.  The SN interaction with this
wind results in a fast outer shock running through the intercloud gas,
and slow shocks running into the dense clouds.  Thus, the
nearly-stationary clouds become engulfed by the outer shock and, for a
certain density contrast and cloud size, can survive in the postshock
gas, eventually encountering the CDS.  If we attribute the fast CS gas
to the effect of shocks running into the engulfed clouds then the
proximity problem may be resolved.  Two options are conceivable,
depending on whether the cloud shock is adiabatic or radiative.  In
the first option, the mechanism which accelerates the unshocked slow
gas in the clouds up to the fast CS gas velocity is presumably the
same as would be the case for a smooth wind in the pre-shock region.
In the case of radiative cloud shocks the fast CS gas may be
identified directly with the shocked cloud gas.  The latter
possibility will be considered in detail in Section 4.

\begin{figure}
\centerline{\hspace{0.0cm}
\psfig{file=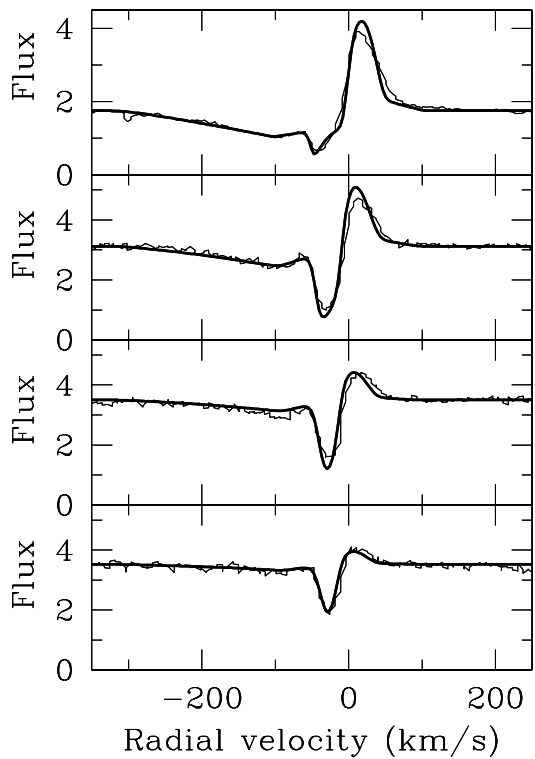,height=12cm,angle=0}
}
\caption{ High resolution Balmer line profiles from H$\alpha$ (top
panel) to H$\delta$ (bottom panel) in SN~1998S on day~42. {\em
Thin line} -- observation, {\em thick line} -- model (Table 1).  }
\label{fig4}
\end{figure}

\begin{table}
  \caption{Parameters of 'best' fit models}
  \bigskip
  \begin{tabular}{ccccccccc}
  \hline
Day  & $R_2$ & $u_{\rm f}$ & $u_{\rm w}$ & $u_{\rm th}$ & $\omega$ 
 & $p$ & $q$ & $n_2$\\ 
\hline 
  &   & \multicolumn{3}{c}{km s$^{-1}$} &  & & & cm$^{-3}$\\ 
\hline
 23  &   1.3 &    500 & 30  &    13   &  0.65  &  6  & 2  & 0.055 \\
 42  &   1.25 &   300 & 30  &    13   &  1    &  2  & 2  & 0.045 \\
\hline 
\end{tabular}
\label{tab1} 
\end{table}

\section{Fast CS gas arising from pre-shock acceleration}

Between the two epochs (23 and 42~d) the photospheric radius expanded
by a factor of $\sim 1.5$ (Fassia et al. 2000).  Yet the above
analysis indicates that, at both epochs, the fast CS gas resided in a
thin shell ($\Delta R/R\approx 0.3$) attached to the photosphere.
Moreover, the gas flow in this shell has a negative velocity gradient,
declining from the maximal velocity to $\approx 30$ km s$^{-1}$ at the
inner edge of the disturbed slow wind.  All this leads us to the
proposition that the fast CS gas is not a pre-existing flow, nor is it
accelerated shortly after the explosion by the radiation from the
shock breakout, but instead is created by the ongoing acceleration of
the slow wind in the immediate vicinity of the photosphere.  The
acceleration might be produced by the SN radiation (Chevalier 1981)
and/or a relativistic particle precursor (Chugai 1997).

\subsection{Radiative acceleration}

The acceleration of an ionised, optically thin wind via Thomson
scattering was invoked to explain the fast CS features observed in
SN~1993J at early epochs (Fransson, Lundqvist \& Chevalier 1996).
Ignoring the initial wind velocity, Thomson scattering can accelerate
the CS gas at radius $r$ up to $v=80E_{49}r_{15}^{-2}$ km s$^{-1}$,
where $E_{49}$ is the time-integrated radiation energy in units of
$10^{49}$ erg, and $r_{15}$ is $r$ in units of $10^{15}$ cm. Fransson
et al. (1996) estimated that $E_{49} \sim 1$ for SN~1993J using
explosion models and calculating the emission from the circumstellar
shock.  For SN~1998S we have no information on the total luminosity
for the early evolution until the estimates for the bolometric
luminosity of Fassia et al. (2000), starting on day 17. Integrating
the estimated bolometric luminosity between days 17 and 42, we obtain
$E_{49}\sim 7$ if we assume that the spectrum is blackbody.  There is
probably considerable uncertainty in this estimate.  We have not
included the early emission from the photosphere and from the
circumstellar interaction, which may lead to an underestimate, but we
have also ignored the effects of UV line blanketing in the blackbody
approximation, thus producing a possible overestimate.  However, the
bolometric luminosity is certainly considerably higher than for SN
1993J, with a likely range of $E_{49} = 5-10$.  Remarkably, in the
SN~II models of Falk \& Arnett (1977) with an extended presupernova
envelope of a size and density similar to those of the SN~1998S inner
shell (Chugai 2001), the radiated energy $E_{49} \approx 10$.

On day 42, $r_{15} \sim 2$ (Fassia et al. 2000) which, with $E_{49} =
5-10$, yields $v\sim 100-200$~km s$^{-1}$ for the CS gas.  This is
faster than the slow wind velocity of $\approx30$ km s$^{-1}$,
suggesting that the observed slow wind may have undergone some
radiative acceleration following the supernova explosion.  Yet these
velocities are insufficient to account for the fast CS shell at this
epoch.  The same is true for day 23 when $r_{15} \sim 1.3$. The value
of $E_{49}$ is even more uncertain at that epoch, probably in the
range $E_{49} = 1-6$, using the results of Fassia et al. (2000).  The
corresponding velocity range for the CS gas is then $v\sim 50-300$~km
s$^{-1}$, again too low for the fast CS gas.  We therefore have to
seek an additional acceleration mechanism to account for the velocity
of the fast CS gas.

\begin{figure}
\centerline{\hspace{-1.cm}
\vspace{-1cm}
\psfig{file=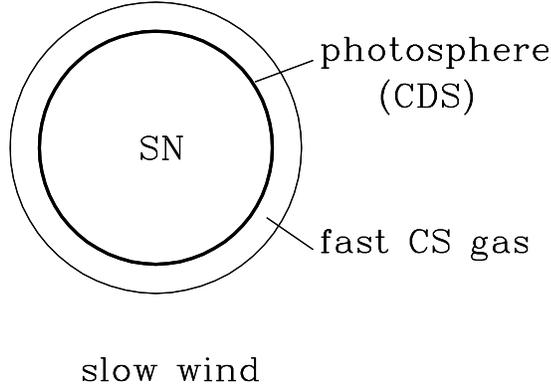,height=9cm,angle=0}
}
\caption{ Illustration of the principal components of the kinematic
model for the fast and slow CS line formation.  }
\label{fig5}
\end{figure}

The radiative force can be enhanced by including the effects of UV
line absorption.  This mechanism was exploited by Castor, Abbott \&
Klein (1975) to account for hot star winds.  It was also invoked by
Chevalier (1981) to explain the fast CS features in SN~1979C via
radiative acceleration in the pre-shock zone, and discussed by
Fransson (1986) who argued that lines could pre-accelerate the gas up
to $\times 4-10$ more efficiently than Thomson scattering.

The radiative force per unit volume due to the line absorption effect
(Castor et al. 1975) is
\begin{equation}
f=\frac{1}{c^2}\left|\frac{d\,v}{d\,r}\right|
\sum F_{\nu}\nu_j[1-\exp (-\tau_j)],
\label{eq:force} 
\end{equation}  

\noindent where $F_{\nu}$ is the monochromatic flux.  $\tau_j$ is the
radial Sobolev optical depth in the line, where $j$ is summed over the
total number of lines with the lines arranged in order of frequency.
From the dependence of the velocity gradient on $f$ we see that line
acceleration is a runaway effect; the gas that has been accelerated is
also the gas that most easily gets further accelerated. This means
that a steeper velocity profile can be created than with pure Thomson
scattering (see also Fransson 1986), thus providing an explanation for
the line profiles in Section~2.  Another mechanism which can steepen
the velocity profile is related to the degradation of the UV radiation
due to photon splitting.  This results in a more efficient
acceleration in the inner layers of the pre-shock zone, where the UV
flux is greater.  Clearly, an adequate treatment of the line effects
in radiative acceleration requires the solution of the radiative
transfer with the statistical equilibrium of ions and level
populations.  This is a formidably complicated task, beyond the scope
of this work.  We shall, instead, restrict ourselves to estimating the
radiative force and then the velocity increase due to the lines
effect.
 
Let $\zeta$ be the ratio of line absorption force $f$ to the Thomson
scattering force $f_{\rm T}=\sigma_{\rm T}n_{\rm e}F/c$ ($F$ is the
total radiation flux integrated over wavelength). Using the equation
of motion, the velocity change of the fast CS gas due to the lines
effect in the pre-shock zone $\Delta R$ may then be written
approximately as
 
\begin{equation}
\Delta u_{\rm f} = \zeta\left(\frac{\Delta R}{R}\right)
\frac{k_{\rm T}L}{4\pi Rv_{\rm s}c},
\label{eq:uprec3} 
\end{equation}  
 
\noindent where $R$ is the radius of the outer shock, $k_{\rm T}=0.3$
cm$^2$ g$^{-1}$ is the Thomson opacity, and $v_{\rm s}$ is the outer
shock wave velocity, which we also identify with the velocity of the
CDS.

To estimate the radiative force due to line absorption we use the list
of $1.5\times10^5$ lines with $\log\,(gf) >-5$ compiled by Ron Eastman
(Blinnikov et al. 1998) and extended here using the R.L.~Kurucz Data Base
available on the web.  Solar composition and a Saha-Boltzmann
distribution are assumed, while the photospheric radiation is treated
as black-body with the flux $F_{\nu}=\pi B_{\nu}(T)$. In Fig.~6 we
show $\zeta=f/f_{\rm T}$ as a function of temperature for densities of
$2\times10^{-17}$, $2\times10^{-16}$ and $2\times10^{-15}$ g
cm$^{-3}$.  (We note that these values are consistent with the lower
limit $n_e \geq 10^{-18}$~cm$^{-3}$ obtained by Fassia et al. (2001)
from observed [O~III] line ratios.)  The velocity gradient in
Eq. (\ref{eq:force}) is taken to be $\Delta v/\Delta R$, with $\Delta
v=400$ km s$^{-1}$ and $\Delta R=3\times10^{14}$ cm.  The results
indicate that $\zeta$ is lower for higher densities. This is due to
line saturation effects [see Eq. (\ref{eq:uprec3})].  The
non-monotonic temperature behaviour is due to the strong temperature
dependence of the ionisation fractions in different ions.  The low
temperature peak at 6000--7000~K is caused primarily by the low
ionisation ions of Fe.
 
On day~23, with the photospheric radius at $1.3\times10^{15}$ cm
(Fassia et al. 2000), the average radius of the fast CS shell is
$\approx1.4\times10^{15}$ cm.  Therefore, a wind linear density
$w\approx10^{16}$ g cm$^{-1}$ would correspond to the density
$5\times10^{-15}$ g cm$^{-3}$.  With the equilibrium temperature
$\approx 1.4\times10^4$~K (Fassia et al. 2000) we have thus $\zeta
\approx 1$ (Fig.~6).  Adopting Eq.(\ref{eq:uprec3})
$R=1.4\times10^{15}$ cm, $v_{\rm s}=6000$ km s$^{-1}$ (Chugai 2001),
luminosity $L=3\times10^{43}$ erg s$^{-1}$ (Fassia et al. 2000), and
$\Delta R/R=0.3$, we obtain from Eq.(\ref{eq:uprec3}) $\Delta u_{\rm
f}\approx 90$ km s$^{-1}$. The integrated acceleration due to Thomson
scattering was estimated above as $50-300$ km s$^{-1}$ for this
particular epoch.  In combination with the lines effect, this yields
$140-390$ km s$^{-1}$ {\it i.e.} the upper limit is about equal to the
observed maximal velocity of the fast CS gas ($\approx 400$ km
s$^{-1}$).  Thus, radiative acceleration is a plausible mechanism for
the acceleration of the fast CS gas. This possibility is enhanced when
we note that the effect of the lines may have been underestimated
since (a) the line list may be incomplete, (b) multiple scattering of
radiation in the UV line forest has been ignored, and (c) the
photospheric spectrum is not strictly black-body.  Also, as has been
shown by Lundqvist \& Fransson (1988) and Fransson et al. (1996), the
wind conditions can be far from thermodynamic equilibrium. Most likely
it is `over-ionised' which would make the wind behave like a gas which
was hotter ({\it cf.} Fig.~6) than the $1.4\times10^4$~K we have
assumed, thereby increasing the radiation force.

The slow-down of the fast CS gas between days 23 and 42 can be
naturally explained using Eq.(\ref{eq:uprec3}). Using information from
Fassia et al. (2000), the ratio $L/Rv_{\rm s}$ decreases by a factor
$\geq 2$ during this period, and unless $\zeta$ increases
significantly during the same period, the line acceleration shows
roughly the right temporal behaviour.

\begin{figure}
\centerline{\hspace{-0.0cm}
\psfig{file=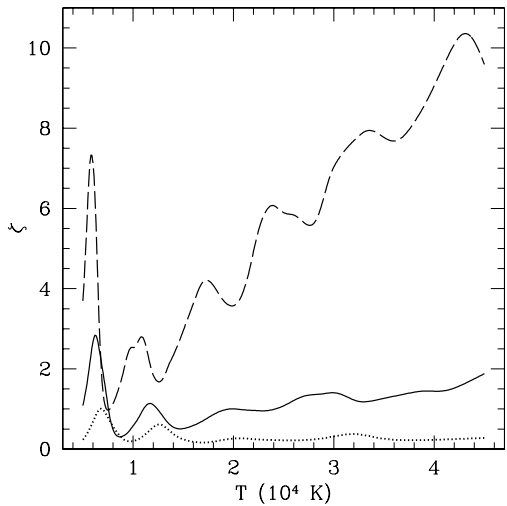,height=7cm,angle=0}
}
\caption{ Ratio of the cumulative line absorption radiation force to
the Thomson scattering radiation force, as a function of the
temperature. Three lines correspond to densities
$\rho=2\times10^{-17}$ g cm$^{-3}$ ({\em dashed}),
$\rho=2\times10^{-16}$ g cm$^{-3}$ ({\em solid}), and
$\rho=2\times10^{-15}$ g cm$^{-3}$ ({\em dotted}).  
}
\label{fig6}
\end{figure}

\subsection{Relativistic particles}

An alternative mechanism for producing pre-shock acceleration is the
dynamical effect of the expected relativistic particle shock precursor
(Kang \& Jones 1997).  Let a fraction $\eta$ of the kinetic
luminosity, $(1/2)wv_{\rm s}^3$, dissipated in the outer shock, be
converted into relativistic particles trapped in the shock wave
region.  In the approximation of constant ejecta velocity, $v_{\rm
s}$, the energy equation with adiabatic losses gives as the total
energy of relativistic particles $E_{\rm r}=(1/4)\eta Rwv_{\rm s}^2$
(where $R$ is the shock wave radius), and the relativistic particle
pressure in the precursor $\Delta p = (1/3)\xi E_{\rm r}/(4\pi
R^2\Delta R)$ (where $\xi$ is the fraction of relativistic particles
in the precursor).  The equation of motion of the pre-shock flow with
velocity $v_1$ in the shock frame, is $\rho \dot{v}_1=-\Delta p/\Delta
R$, and so we have as the velocity of the pre-shock gas in the lab
frame
 
\begin{equation}
u_{\rm f}= \frac{1}{12}\left(\frac{R}{\Delta R}\right)\xi\eta v_{\rm s}. 
\label{eq:vfcr}
\end{equation}   
 
\noindent Setting $\Delta R/R=0.3$ and $v_{\rm s}=6000$ km s$^{-1}$,
to obtain a precursor velocity of $u_{\rm f}=500$ km s$^{-1}$ we need
$\xi\eta\approx0.3$. Given comparable fractions of relativistic
particle energy in the post-shock and pre-shock regions (Kang \& Jones
1997), we may adopt $\xi \approx 0.5$.  Thus, to accelerate pre-shock
matter up to 500 km s$^{-1}$, the conversion efficiency of shock
energy into relativistic particles must be $\eta\approx 0.6$.  While
this value is close to the absolute upper limit, it is nevertheless
plausible given the high efficiency of particle acceleration in high
Mach number astrophysical shocks (Blandford 1984).  Thus, from energy
considerations alone, we cannot rule out the possibility that the
relativistic particle precursor may significantly contribute to the
acceleration of the wind in SN~1998S.

\section{Fast CS gas arising from shocked wind clumps}

We now consider a radically different mechanism for producing the fast
CS gas.  This has already been mentioned in Section 2.  It essentially
exploits the possibility that the wind may have a clumpy structure.
In this scenario, the fast CS gas is identified with
radiatively-cooled shocked gas in wind clouds engulfed by the
intercloud shock.  This may, in principle, reproduce both the
characteristic velocity of the fast CS gas and resolve the photosphere
proximity problem.  Moreover, it may also produce a negative velocity
gradient.  Since the postshock density $\rho_{\rm ps}$ increases from
the outer shock jump towards the contact discontinuity (Chevalier
1982), it means that the cloud shock velocity $v_{\rm c}\approx v_{\rm
s}(\rho_{\rm ps}/\rho_{\rm c})^{-1/2}$ also increases inwardly.  Thus,
the negative velocity gradient may well be mimicked by the ensemble of
cloud shocks.

A specific feature of the proposed mechanism is a `stripping flow'
produced by the mass loss from the shocked cloud due to a
Kelvin-Helmholtz instability (Klein, McKee \& Colella 1991).  The
stripping flow is characterized by a broad spectrum of fragment sizes
and velocities with the smaller fragments acquiring the higher
velocities.  Tiny, high velocity fragments are subjected to efficient
mixing with the hot outer shock gas, which results in their heating
and ionisation.  Therefore, we expect that the high velocity stripping
flow should be revealed particularly by the presence of high
ionisation lines.  In fact HST UV spectra do show high velocity
(several $\times 10^3$ km s$^{-1}$) blue wings in C~IV and Si~IV
resonance lines (Lundqvist et al., in preparation), and this may well
be the consequence of such a stripping flow.

We now test the clumpy wind model against particular observational
requirements, particularly on day 23.  The cloud/intercloud density
ratio $\chi$ must be consistent with the cloud shock wave velocity,
$v_{\rm c}=2v_{\rm s}\chi^{-1/2}$ (the factor of two is from the
factor of four density jump at the intercloud shock).  With $v_{\rm
c}\approx 500$ km s$^{-1}$ and a CDS velocity at this epoch of about
$v_{\rm s}\approx 6000$ km s$^{-1}$ we thus obtain $\chi\approx 600$.
To meet the requirement of proximity to the photosphere, the lifetime,
$t_{\rm c}=2a/v_{\rm c}$, of a cloud of radius $a$ must be equal to or
greater than the time, $\Delta R/v$, which it takes to cross the
intercloud post-shock layer ($\Delta R$ is the thickness of this
layer).  This leads to the constraint $a/\Delta R\geq 1/\sqrt
{\chi}\approx 0.04$.  We consider the case where the ratio is close to
this lower limit, {\it viz.} $a/\Delta R\approx 0.05$, in order to
maximize the number of shocked clouds and so be consistent with the
absence of strong fluctuations in the line profile.  Combining this
ratio with $\Delta R/R \approx 0.2$ and the value of the photospheric
radius at this epoch, $R\approx 1.3\times 10^{15}$ cm (Fassia et
al. 2000), we get a cloud radius $a\approx 10^{13}$ cm.  The filling
factor of the clouds ($f$) may be estimated from the condition that
the covering factor of the shocked clouds is about unity, {\it i.e.}
$(3/4) f(\Delta R/a)\approx 1$.  From this, we obtain $f\approx 0.07$.
Adopting all the cloud parameters derived above, we find that the
number of clouds in the outer postshock layer is $N=3f(R/a)^3(\Delta
R/R)\approx 9\times10^4$.
 
We now demonstrate that, given the above choice of cloud parameters
($a$, $\chi$, and $f$), the cloud shocks are essentially radiative.
The condition that the cooling time is less than the time for the
cloud shock wave propagation is $a/v_{\rm c} > m_{\rm p}v_{\rm
c}^2/(32\Lambda n_{\rm c})$, where $m_{\rm p}$ is the proton mass, and
$\Lambda \approx 2\times10^{-23}$ erg s$^{-1}$ cm$^3$ is the cooling
function.  This leads to the requirement $an_{\rm c}>3\times10^{20}$
cm$^{-2}$, where $v_{\rm c}=500$ km s$^{-1}$ is assumed.  Given an
average wind density $n=w/(4\pi R^2 m_{\rm p})\approx 2\times10^8$
cm$^{-3}$, $\chi=600$, and $f=0.07$, we obtain a cloud density of
$n_{\rm c}\approx 3\times 10^9$ cm$^{-3}$.  Therefore, from the above
lower limit for the cloud column density, $an_{\rm c}$, we find that
for the cloud shocks to be radiative, $a>2\times10^{11}$ cm.  This is
consistent with the value $a\approx 10^{13}$ cm derived above using the
observed values for $v_{\rm c}$, $v_{\rm s}$ and $R$ at 23~days.
 
Finally, the adopted set of cloud parameters must be consistent with
the day~23 observational upper limit for the net H$\alpha$ luminosity
$L({\rm H}\alpha)<10^{39}$ erg s$^{-1}$ (Fassia et al. 2001).  To
check this, we use the simple expression for the cumulative H$\alpha$
luminosity of the shocked clouds

\begin{equation}
L({\rm H}\alpha)=\frac{3}{8}\psi wv_{\rm c}^3 \left(\frac{\Delta R}{a}
\right)
\frac{\chi f}{1+\chi f},
\label{eq:lum}
\end{equation} 
  
\noindent where $\psi$ is the efficiency of H$\alpha$ emission, of the
order of $10^{-2}$ (Chevalier \& Fransson 1994).  Substituting in
Eq. (\ref{eq:lum}) $w=10^{16}$ g cm$^{-1}$, $v_{\rm c}=500$ km
s$^{-1}$, $\psi=0.01$, $a/\Delta R=0.05$, one obtains $L({\rm
H}\alpha)\leq 10^{38}$ erg s$^{-1}$, in accord with the observational
upper limit.

Thus, for the cloud shock scenario an appropriate set of cloud
parameters might be $\chi\approx 600$, $f\approx 0.1$, $a\approx
10^{13}$~cm, with the average linear wind density $w\approx 10^{16}$ g
cm$^{-1}$.  Note, this model can account for the observed slow down
with time of the average cloud shock velocity ($v_{\rm c}\approx
v_{\rm s}/\sqrt{\chi}$), if the density contrast $\chi$ increases with
radius between days 23 and 42 and/or $v_{\rm s}$ decreases with time,
both being plausible.

Summing up, the above order of magnitude analysis shows that the
shocked wind cloud model for the fast CS component may be matched to
the bulk of the empirical constraints for a reasonable set of
parameters.  While the proposed model is quite crude, it does appear
to offer a useful alternative explanation for the fast CS gas
phenomenon, including the high velocity blue wings of the fast CS
absorption profiles in high ionisation UV lines.

\section{Discussion}

The principal finding of this paper is that the fast CS gas observed
in Balmer lines in SN~1998S during the era day~23--42 has a negative
velocity gradient flow confined to a narrow layer above, and attached
to, the photosphere.  The same should be true for the fast CS gas
observed in He~I (Bowen et al. 2000, Fassia et al 2001) and Mg~II 2800
\AA\ (Bowen et al. 2000).  We argue that the fast CS gas is the result
of the acceleration of the slow wind in the immediate vicinity of the
SN photosphere, and show that acceleration driven by radiation is a
plausible mechanism.  Thomson scattering alone, while significant, is
insufficient to account for the observed fast CS gas velocity.
However, when we include the additional effects of UV line absorption,
we can obtain the observed velocity.  Moreover, this is a conservative
conclusion given the that the line list may be incomplete, that we
ignore multiple scattering in UV line forest, and that time-dependent
and non-thermodynamic equilibrium effects are not included.  We also
demonstrate that the a relativistic particle precursor is able to
accelerate pre-shock matter up to the observed fast CS gas velocity.

The fast CS gas can also originate in a completely different way; it
could be the result of radiative shocks propagating into dense wind
clumps together with the effects of a stripping flow.  The stripping
flow aspect of the clumpy wind model predicts the presence of very
fast, highly ionised CS gas.  This seems to be supported by the
observation of high velocity blue absorption wings (several $\times
10^3$ km s$^{-1}$) in C~IV and Si~IV (Lundqvist et al., in
preparation).
 
An intriguing fact, is that the narrow line width of the unaccelerated
wind ($\approx 15$ km s$^{-1}$) indicates a rather moderate wind
temperature $\sim 10^4$ K.  This is quite unlike the hot wind ($\sim
10^5$ K) at early epochs ($t\sim 1$ month) anticipated by the model of
Lundqvist \& Fransson (1988) for SN~1979C with comparable wind
density, as well as for the somewhat less dense wind around SN~1993J
(Fransson et al. 1996).  This disagreement may indicate that either
the outer shock is essentially radiative so that the Comptonization
responsible for the bulk of the UV radiation is inefficient, or that
the wind is clumpy with a resulting low equilibrium temperature in the
pre-shock wind clumps.  We favour the second option, because a
radiative outer shock would produce even higher UV luminosity than
that of Comptonized radiation from an adiabatic shock.  We note that a
clumpy-wind scenario was also introduced for SN~1993J by Fransson et
al. (1996) to resolve a similar controversy between the theoretical
high wind ionisation and a lower ionisation implied by the CS spectral
lines.

In Fassia et al. (2001) we suggested that the fast CS gas in SN~1998S
resulted from the progenitor going through an RSG phase and then a BSG
phase shortly before it exploded.  `Mid-CSM' and `Outer-CSM' zones
were defined corresponding, respectively, to a tenuous, fast CSM wind
from the BSG, and a dense, slow CSM wind from the RSG wind.  The BSG
wind would have `piled up' against the inner boundary of the RSG wind,
producing a negative velocity gradient.  The SN shock would first
engulf the faster more tenuous part of the BSG wind thus accounting
for the observed decrease in the absorption blueshift with time.  The
abrupt appearance of the strong broad lines by about day~45 would be
due to the impact of the ejecta with the denser RSG wind.  Thus,
qualitatively at least, this model can account for much of the
observed spectral behaviour.  However, an important result of the
spectral line modelling described in Section~2 is that, in terms of
the photospheric radius, the radius, $R_2$, of the interface between
the accelerated (fast) and unaccelerated (slow) wind remains about
constant between days~23 and 42 (Table~2).  Yet, in the RSG/BSG wind
explanation, $R_2$ decreases from 2.1 on day~23 to 1.3 on day~42.  But
Model~4 (Table~1 and Fig.~2) shows that changing $R_2$ from 1.3 to just
1.5 produces a much poorer reproduction of the observed profiles.
This appears to present a difficulty for the BSG/RSG wind scenario.
Moreover, the acceleration model can also account for the sudden
appearance of the strong, broad lines simply as a result of the CDS
becoming optically thin around this epoch (Chugai 2001).  In other
words, there may not be a need for a division of the wind into
`Mid-CSM' and `Outer-CSM' components.  All this leads us to favour the
proposition that the fast CS gas is not a pre-existing flow, but
instead is created by the ongoing acceleration of the slow wind in the
immediate vicinity of the photosphere.  The question of whether or not
the BSG/RSG explanation can survive quantitative modelling is beyond
the scope of this paper.
 
It is instructive to review briefly the other known cases of type~IIn
and type~IIL supernovae which have high velocity CS spectral features.
The first observational evidence for a phenomenon of this kind was
provided by the IUE spectra of the type~IIL SN~1979C.  This revealed
UV lines of C~IV and Si~IV indicating a gas velocity in the range
$1000-4000$ km s$^{-1}$ (Panagia et al. 1980).  Bearing in mind the
photometric and spectral similarities between SN~1998S and SN~1979C,
the fast (CS) material in SN~1979C may also be associated with shocked
wind clumps.  Originally, the UV lines in SN~1979C were attributed to
pre-shock radiative acceleration of the slow wind (Chevalier 1981), or
to emission from X-ray heated ejecta (Fransson 1984).

Another case is the type~IIb SN~1993J where H$\alpha$ P~Cygni profiles
suggested a gas velocity of $\approx 1000$ km s$^{-1}$ at 2--3 days
post-explosion (Cumming et al. 1994). A similar velocity was indicated
by the N~V~1240 \AA\ emission line (Sonneborn et al.  1994).  Fransson
et al. (1996) suggested radiative acceleration of the slow wind as a
mechanism for the origin of this fast CS gas, but noted that Thomson
scattering could be insufficient. It would be interesting to determine
whether or not line absorption radiation pressure could provide the
required enhancement of the radiative force in this case.  Or do we
need alternative mechanisms such as a cosmic ray precursor or shocked
wind clouds?

In the type~IIn SN~1984E, spectral features are attributed to a superwind.
Velocities of $<1000$ km s$^{-1}$ are indicated by the emission
components of Balmer lines, while the blue edge of the H$\alpha$
absorption indicates a velocity as high as 3000 km s$^{-1}$ (Dopita et
al. 1984), i.e. substantially higher than in the case of SN~1998S.
The low signal-to-noise ratio of the SN~1984E data leaves some doubts
about the precise interpretation of the line profile; in any event a
velocity of 3000 km s$^{-1}$ at the observed epochs is probably beyond
the capability of radiative acceleration.
 
The type~IIn SN~1994W exhibited P~Cygni lines which indicated CS gas
velocities of $10^3$ km s$^{-1}$.  At first sight, therefore, we might
take this SN to be similar to SN~1998S.  However, the amazing
persistence of the absorption line velocities for about two months
(Sollerman, Cumming \& Lundqvist 1998) is quite unlike the velocity
evolution observed in the fast CS gas features of SN~1998S.  A
pre-existing CS shell (wind) with a constant velocity of $\approx
1000$ km s$^{-1}$ is the more likely explanation for the fast CS gas
around SN~1994W (Sollerman et al. 1998).

It may well be that in different SNe~IIn we see two kinds of fast CS
matter.  One type is the result of acceleration of the slow progenitor
wind at the ejecta-wind interface (e.g. SN~1998S), while the other
type is produced by the progenitor before it explodes (e.g. SN~1994W).
The origin of such different modes of heavy mass loss in the final
stages of a star's life remains quite uncertain, with the conceivable
involvement of stars from both the low mass ($\sim 10$~M$_{\odot}$)
and high mass ($>20$~M$_{\odot}$) ends of the massive star range.  By
studying a sufficiently large sample of SNe~IIn using high-resolution
spectroscopy together with other observations in the X-ray, optical
and radio bands, we shall acquire clues about the complicated physics
of the ejecta-wind interaction, the origin of the dense CS matter, and
the pre-SN~IIn genesis.

\section*{Acknowledgments}

The research made by one of us (NC) was partially supported by RFBR 
grant 01-02-16295. We also acknowledge support from The Royal Swedish Academy 
of Sciences, The Wenner-Gren Foundations and the Swedish Research Council. 
We also acknowledge support from the UK Particle Physics and Astrophysics 
Research Council through grant PPA/G/S/1997/00266. The work of SB and ES was 
supported by RFBR grant 99-02-16205.


\begin{thebibliography}{}


\bibitem{}
Blandford R.D., 1986, in Magnetospheric phenomena in astrophysics, 
 ed. R.I.~Epstein \& W.C.~Feldman (New York: AIP), 1

\bibitem{}
Blinnikov S.I., Eastman R., Bartunov O.S., Popolitov V.A., Woosley S.E., 
 1998, ApJ, 496, 454

\bibitem{}
Bowen D.V., Roth K.C., Meyer D.M., Blades J.C., 2000, ApJ, 536, 225

\bibitem{}
Castor J.I., Abbott D.C., Klein R.I., 1975, ApJ, 195, 157

\bibitem{}
Chevalier R.A., 1981, ApJ, 251, 259

\bibitem{}
Chevalier R.A., 1982, ApJ, 259, 302

\bibitem{}
Chevalier R.A., Fransson C., 1985, in Supernovae as distance 
 indicators, ed. N.~Bartel (Berlin: Springer), 123

\bibitem{} 
Chevalier R.A., Fransson C., 1994, ApJ, 420, 268

\bibitem{}
Chevalier R.A., Blondin J.M., 1995, ApJ, 444, 312

\bibitem{}
Chugai N.N.,1993, ApJ, 414, L101

\bibitem{}
Chugai N.N., 1997, Astrophys. Space Sci., 252, 225

\bibitem{}
Chugai N.N., 2001, MNRAS, submitted

\bibitem{}
Cid Fernandes R.C., Terlevich R., 1994, in Tenorio-Tagle G. ed.
 Violent Star Formation. From S Doradus to QSO. CUP, Cambridge p. 365

\bibitem{}
Cumming R.J., Meikle P., Walton N., Lundqvist P., 1994,
 in Circumstellar media in the late stages of stellar evolution,
 ed. R.E.S.~Clegg, I.R.~Stevens and W.P.S.~Meikle (Cambridge: Univ. Press),
 192

\bibitem{}
Dopita M.A., Evans R., Cohen M, Schwartz R.D., 1984, ApJ, 287, L69
 
\bibitem{}
Falk S.W., Arnett W.D., 1977, ApJS, 33, 515

\bibitem{}
Fassia A., Meikle W.P.S., Vacca W.D. et al., 2000, MNRAS 318, 1093

\bibitem{}
Fassia A., Meikle W.P.S., Chugai N.N. et al., 2001, MNRAS, in press

\bibitem{}
Filippenko A.V., 1997, ARA\&A, 35, 309

\bibitem{}
Fransson C., 1984, A\&A, 132, 115

\bibitem{}
Fransson C., 1986, Highlights Astron., 7, 611

\bibitem{}
Fransson C., Lundqvist P., Chevalier R.A., 1996, ApJ, 461, 993

\bibitem{}
Grasberg E.K., Nadyozhin D.K., 1987, Sov. Astron., 31, 629

\bibitem{}
Kang H., Jones T.W., 1997, ApJ, 476, 875

\bibitem{}

Klein R.I., McKee C.F., Colella P., 1991, in Supernovae, ed. S.E.~Woosley 
 (New York: Springer), 696

\bibitem{}
Leonard D.C., Filippenko A.V., Barth A.J., Matheson T., 2000, ApJ, 538, 239

\bibitem{} 
Lentz E.J. et al., 2001, ApJ, 547, 406

\bibitem{}
Li W.D., Wan Z., 1998, IAU Circ., No. 6829

\bibitem{}
Liu Q.-Z., Hu J.-Y., Hang H.-R., Qiu Y.-L., Zhu Z.-X., Qiao Q.-Y., 
 2000, A\&AS, 144, 219

\bibitem{}
Lundqvist P., Fransson C., 1988, A\&A, 192, 221

\bibitem{}
Nadyozhin D.K., 1981, Preprint ITEP No. 1

\bibitem{}
Nadyozhin D.K., 1985, Astrophys. Space. Sci., 112, 225

\bibitem{}
Olofsson H., 1994,
 in Circumstellar media in the late stages of stellar evolution,
 ed. R.E.S.~Clegg, I.R.~Stevens and W.P.S.~Meikle (Cambridge: Univ. Press),
 246

\bibitem{}
Panagia N., Vettolani G., Boksenberg A. et al., 1980, MNRAS, 192, 861

\bibitem{}
Pooley D., et al. 2001, astro-ph/0101196

\bibitem{}
Sollerman J., Cumming R.J., Lundqvist P., 1998, ApJ, 493, 933

\bibitem{}
Sonneborn G., Pascual P.R., Wamsteker W., Fransson C., 1994,
 in Circumstellar media in the late stages of stellar evolution,
 ed. R.E.S.~Clegg, I.R.~Stevens and W.P.S.~Meikle (Cambridge: Univ. Press),
 198

\end{thebibliography}
\end{document}